\documentclass{article}
\usepackage{graphics}

\begin{document}
\title{Exfoliation of singlewall carbon nanotubes
in aqueous surfactant suspensions : a Raman study}

\author{N. Izard$^{1,2}$, D. Riehl$^1$, E. Anglaret$^2$\\
1-Centre Technique d'Arcueil, DGA, Arcueil, France\\
2-Groupe de Dynamique des Phases Condens\'ees, UMR CNRS 5581,\\
Universit\'e Montpellier II, Montpellier, France\\}

\date{\today}
\maketitle
\vskip 1.5cm

\begin{abstract}
Aqueous suspensions of {\it bundles} and {\it individual} singlewall carbon nanotubes
were prepared with the help of surfactants. We study the changes in the Raman spectra
of the suspensions with respect to powders, and of exfoliated tubes with respect to bundles.
The radial breathing modes (RBM) upshift in suspensions because
of the internal pressure of the liquid. By contrast, no shift is observed in the RBM
spectra after exfoliation in the suspensions. However, we demonstrate a selectivity of
the exfoliation process for tubes of small diameters.
\end{abstract}

\vskip 1cm

\\

{\bf I. INTRODUCTION}

Raman spectroscopy is certainly the most widely used technique to study single
wall carbon nanotubes (SWNT) \cite{drerev,jlsrev}. Raman scattering from SWNT is
a resonant phenomenom and therefore the signal is intense and easy to measure.
The popularity of the technique is also due to the availability of simple models
to interpret the Raman data which provide useful information not only on the vibrational
properties but also on the stucture and electronic properties of the nanotubes. Note
however that the analysis of some specific features of the spectra for small bundles
and individual tubes is still in debate.\\

Measurement of the radial breathing modes (RBM) is the most direct
and efficient way to estimate the diameter distribution of SWNT in
a sample. Calculations using a force constant approach early
predicted that the frequency of the RBM for an {\it isolated} tube
is proportional to the inverse of the tube diameter
\cite{drerev,jlsrev,bandow,rols}. Note that this linear law was
never really checked by experiments because series of isolated
tubes of various diameter are not readily available. Indeed, most
of the available samples contain bundles of some tens to some hundreds of nanotubes.
Calculations were therefore extended to bundles, in a first step to infinite ones, using tight
binding \cite{ven,hen1}, or force constant \cite{rols,rah2} models. The frequency of the RBM was
found to be upshifted of 5 to 15$\%$ depending on the model. An experimental check of these calculations
was performed by Rols {\it et al} on a series of samples featured by different diameter distributions
\cite{rols} and a good agreement was found between the mean diameter estimated from X-ray diffraction
measurements and from RBM frequencies. The periodic photoselective resonance of SWNT in a polydisperse
sample was qualitatively well described by combining {\it ab initio} calculations and evaluations of the
resonance cross-sections of the nanotubes \cite{mil}. Therefore, a quantitative interpretation of the RBM
spectra for bundles of SWNT can be achieved providing tube-tube interactions within the bundles are well
taken into account \cite{rols,mil}. On the other hand, numerous Raman studies were performed on isolated
tubes \cite{dreisoles,duesberg}, but only a few experimental studies addressed the changes in the Raman
spectra when the nanotubes are exfoliated from the bundles. Rao {\it et al} compared the Raman signatures
of dry powders (containing essentially bundled tubes) and solutions of individual functionalized tubes
\cite{rao}. They found an unexpected upshift of the RBM frequency for the solubilized tubes. Recently,
Strano {\it et al} reported changes in the relative intensities of the RBM bands as a function of the
aggregation state of the nanotubes in suspensions, floculates, powders \cite{strano}... In these two
studies, changes in the RBM spectra were assigned to changes in the Raman resonance conditions, due
to changes in the electronic properties of the nanotubes as a function of the aggregation state.
However, it was shown by Wood and Wagner that the internal pressure of the media may induce a shift
of some Raman modes, especially those involving radial motions of the carbon atoms \cite{wag1}. Therefore,
a relevant comparison of the Raman spectra for bundles and individual tubes can only be made when the
nanotubes are dispersed in the same fluid.\\

In this paper, we study aqueous suspensions of {\it individual}
SWNT (SIS) and we focus on the changes in the Raman spectra from
powders to suspensions and from suspensions of bundles (SB) to
SIS. The preparation of the samples, their characterization by
several techniques and the demonstration of exfoliation in SIS is
presented in section II. The raw powders, the SB and the SIS made
from the same samples were studied by Raman spectroscopy using
various laser lines. The experimental results are presented and
discussed in section III. In a first part (subsection III-A), we
focus on the changes observed in the RBM range for the suspensions
with respect to powders. We show that the internal pressure of the
liquid is responsible for an upshift of the RBM in the
suspensions. In a second part (subsection III-B), we study the
changes in the Raman spectra after exfoliation. No further shift
can be observed in the RBM range between SB and SIS. By contrast,
the relative intensities of the RBM bands change. We consider two
hypothesis to explain these changes : i) changes in the electronic
properties due to exfoliation and ii) selectivity of the
exfoliation process in favor of
the smallest diameters.\\

{\bf II. EXPERIMENTAL}\\

{\bf A. Preparation and characterisation of the samples}\\

In this study, we worked on three different kinds of single wall carbon nanotubes : raw and purified
samples prepared by the electric arc (EA) technique (from Nanoledge, Inc and MER, Inc, respectively)
and purified samples prepared by the HiPCO technique (from CNI, Inc). As far as RBM are concerned,
the Raman results are very close for raw and purified EA samples and we will present only those for
raw samples in this paper. The powders were extensively characterized by X-ray diffraction, scanning
electron microscopy (SEM), high resolution transmission electron microscopy (HRTEM), Raman spectroscopy,
optical spectroscopy, in order to check their homogeneity, purity, and to estimate the distribution of
diameter for the tubes and the bundles. A selection of the results is presented in reference
\cite{procKir}. The SWNT diameter is typically between 1.2 and 1.5 nm for the EA samples and between 0.7
and 1.3 nm for the HiPCO samples. The bundle diameter  is typically between 20 and 30 nm for the EA samples
and it is significantly smaller (of the order of 10 nm) for the HiPCO samples.\\

Suspensions were prepared from the three batches of samples,
following the procedure described by Vigolo {\it et al} in
reference \cite{vig} for the bundles and that described by
O'Connell {\it et al} in
reference \cite{oco} for individual tubes. For SB, we prepared suspensions with about 200 mg.l$^{-1}$
of nanotubes. For all samples, we prepared the suspensions using sodium dodecyl sulfate (SDS) which is
an ionic surfactant, also used in references \cite{oco,bac}. We also prepared suspensions of bundles for
the EA samples using Triton X100$^{TM}$ (TX100), which is a non ionic surfactant. No changes in the RBM
spectra were observed when the surfactant changes for raw EA and HiPCO samples. By contrast, some changes
were observed in the spectra of the tangential modes for purified EA samples and attributed to charge
transfers between surfactant and tubes. However, this falls out of the scope of this paper and this will
be adressed in details elsewhere. Note finally that some suspensions were prepared using
D$_2$O instead of H$_2$O in order to prevent from optical absorption in the near infrared. However,
in our backscattering geometry, the beam can be focalized on the front face of the cell so that absorption
from the dispersant is considerably reduced. As a matter of fact, we experimentally checked that the
Raman profiles in the RBM and TM ranges are absolutely not affected by the nature of the isotope.\\

The preparation of individual nanotubes was controled by HRTEM.
A dialysis of the suspensions in SDS with a pH 8.5 tris(hydroxymethyl)aminomethan buffer solution
was first achieved to remove the surfactant.
The samples were then diluted in ethanol, and a few drops of this latter
solution were deposited on HRTEM grids where the solvent was evaporated. Typical images of bundles
(from the powder) and individual tubes are presented in figures 1a and 1b, respectively. Imaging the
 bundles is rather straightforward. Well-crystallized bundles are observed, as stated by well-defined
  interference fringes. A few individual tubes and a few carbon and catalyst nanoparticles can be observed
  as well. Imaging individual tubes is more difficult. No bundles could be observed in the samples. Another
   important result is that at ''low'' magnification (typically that of picture 1a), only a few individual
   tubes can be observed, those which peep out of the carbon skin deposited on the grids (not shown). For
   most of these individual nanotubes, one can observe one of the extremities contrarily to the bundles
   imaged in fig. 1a where both extremities are embedded in other bundles or in the carbon skin. Therefore,
    we confirm that the individual tubes are probably significantly shortened by the sonication and
    centrifugation treatments, as already claimed in references \cite{oco,bac,har}. However, it would be
     presomptuous to try to estimate the distribution of nanotube lengths from the pictures since at
     higher magnification (fig. 1b), most of the imaged nanotubes are lying on the carbon skin and their
     extremities can not be identified accurately.\\

Other evidences of the exfoliation of the tubes were obtained. First, for the two samples, SIS present
 a strong photoluminescence (PL) in the near infra-red, while powders and SB do not (figure 2). The PL
 spectra are in good agreement with those reported in the literature for similar samples \cite{bac,leb}.
  Near infrared PL is a direct evidence of the exfoliation of the tubes in the suspensions (we remind that
   the contacts between metallic and semi-conducting tubes in the bundles quench the fluorescence \cite{oco}).
   Note that in the specific case of HiPCO samples, fluorescence is also observed from suspended bundles
   (fig. 2b), but with a weaker intensity. This is an indication that HiPCO samples contain individual
   tubes or thin bundles, which is not surprising since the typical diameter of the original bundles in
   the powders for HiPCO samples is much smaller than for EA samples \cite{procKir}. Furthermore,
   it is likely that the sonication process favors the exfoliation of small tubes, especially if they are
   located at the surface of polydisperse bundles, because of weaker van der Waals interactions with neighbour
   tubes due to their strong curvature angle and small
diameter. We will give below (section III-B) further evidences of
this selectivity. Finally, we also got other evidences of the
preparation of individual tubes from photon correlation
spectroscopy experiments \cite{bad} and from optical limiting
measurements \cite{iza-ol}.
These results will be presented in details elsewhere.\\

{\bf B. Raman scattering experiments}\\

The Raman experiments were carried out for the three batches of samples under three forms (powder, SB and SIS)
at three laser energies (2.41 eV and 1.92 eV from an Ar-Kr ion laser, and 1.16 eV from a Nd-YAG laser).
Measurements
were achieved in both slightly focused (spot diameter of the order of some hundreds micrometers) and
tightly focused (spot diameter of the order of ten
micrometers) configurations and gave the same results. The laser power was below 100 W.cm$^{-2}$ for the
study of powders to prevent the samples from heating. For the suspensions, we checked that the laser power
density can be increased by
more than a factor of ten without any changes in the profile of the Raman spectra, {\it i.e.} without any
significant heating of the samples. This is due to the much larger thermal conductivity of the liquids with
 respect to the air.\\

{\bf III. RAMAN RESULTS AND INTERPRETATION}\\

{\bf A. Effect of environment}\\

Figure 3 presents Raman spectra in a low-frequency range excited with three different laser lines. All the
peaks measured in this low-frequency range are assigned to radial breathing modes (RBM). The relation between
RBM frequency and  tube diameter was calculated using a force constant model and was checked experimentally
for powders of bundled tubes \cite{rols} :

    $$\nu (cm^{-1})=\frac{224}{d (nm)} +14$$    (1)

On the other hand, it is well known that Raman scattering is a
resonant phenomenom for SWNT. The resonance conditions directly
relate to allowed optical transitions (AOT) between pairs of van
Hove singularities, which are commonly reported as a function of
tube diameter on the so-called ''Kataura plot''
\cite{drerev,jlsrev,kat,hert}. The RBM spectra for EA powders
(fig. 3a) can be well interpreted using equation (1) and the
Kataura plot. The spectra at 2.41 and 1.92 eV correspond to
resonance on semi-conducting and metallic tubes, respectively,
with diameters between 1.2 and 1.5 nm. The spectrum at 1.16 eV
corresponds to resonance on semi-conducting tubes between 1.3 and
1.5 nm. The interpretation of the spectra for tubes of small
diameter, as in the case of HiPCO samples (fig. 3b), must be made
with more care since i) equation (1) was experimentally checked
only for tube diameters above 1 nm and ii) corrections to the
simple zone folding model are necessary to calculate the density
of states, and therefore the AOT, of small tubes
\cite{hert,stra2}. The RBM observed at larger frequencies, between
200 and 300 cm$^{-1}$, are assigned to smaller tubes of diameter
between 0.7 and 1.2 nm, with resonance on metallic tubes at 2.41
 eV and on semiconducting tubes at 1.92 and 1.16 eV. No RBM is measured below 185 cm$^{-1}$ which confirms
 that there are essentially no tubes of diameter above 1.3 nm in HiPCO samples.\\

The most striking feature of figure 3 is the systematic upshift of the RBM bunch when SWNT are dispersed
in suspensions. The upshifts can be as high as 7 cm$^{-1}$ at low frequencies (fig. 3a) and are more modest
 at high frequencies (fig. 3b). In order to estimate more precisely the diameter dependence of the shift,
 all spectra were fitted with a set of lorentzians. The fitting procedure was the following : i) for the
  powder spectra, the number of lorentzians was first adjusted in order to fit well each secondary maxima.
  A line was added if one of the linewidths (FWHM) was larger than 15 cm$^{-1}$, a line was removed if its
  integrated intensity was below 2$\%$ of that of the whole bunch, ii) the fit of the spectra for the
  suspensions was achieved with the same number of lines than the powder, and started with the same fitting
   parameters. A line was removed only if its integrated intensity was below 2$\%$ of that of the whole
   bunch, which occured occasionally. {\it Systematic} upshifts of the lines were observed. On the other hand,
   good fits could {\it never} be achieved by fixing the line frequencies and varying the intensities only.\\

 The shift of each RBM peak from powder to suspension is reported in figure 4 for each sample and each laser
line. The systematic upshift is incontestable, as well as its
frequency dependence. For a given RBM frequency, the best linear
fit of the data is very close for SB and SIS, for EA and HiPCO
samples, for measurements at 2.41 eV and 1.92 eV (not shown). Only
for the results at 1.16 eV, the shift was found to be
significantly smaller (triangles in figure 4). However, one can
observe a rather important dispersion of the data. This dispersion
can be due to different bundle sizes for SB or different
organisations of the surfactant as a function of chiral angle.
This can also be due to changes in the electronic properties of
the tubes as a function of chemical environment or bundle size,
especially for SIS. If this occurs, part of the measured shift may
be assigned to a change of the nature of the resonant tube. These
dispersion effects may explain why the apparent shift at 1.16 eV
is smaller than for the other laser lines. Given the small number
of data at 1.16 eV, it would be presomptuous to go further in the
interpretation. Finally the best fit of the whole data (solid line
in figure 4) gives :

    $$\Delta\nu (cm^{-1})=12.2-3.7.10^{-2} \nu (cm^{-1})$$  (2)

Shifts of the same order of magnitude were already reported i) by Rao {\it et al} and assigned to
an exfoliation of fonctionalized and shortened tubes in CS$_2$ solutions (figure 1 in \cite{rao}), and ii) by Lebedkin {\it et al} but not interpreted (figure 3 in \cite{leb}). We claim that these upshifts {\it are not} due to exfoliation of the tubes since they are observed both for suspensions of bundles {\it and} individual tubes. For the same reason, we rule out the assignment of the RBM shift to bundle thickening, which was proposed by Kukovecz {\it et al} to explain Raman results on functionalized nanotubes \cite{kuko}. On the other hand, a systematic upshift of the RBM can not be explained by a systematic offset of the AOT due to changes in the chemical environment, since the apparent RBM shift for each peak of the bunch could be positive or negative depending on the laser line and the resonant tubes.

We claim that the RBM upshift must rather be assigned to changes
in the interactions between nanotubes and its environment. Indeed, it is
instructive to compare the shifts measured in figure 4 to those
measured under hydrostatic pressure \cite{ven,san}. An upshift of
a few cm$^{-1}$ was observed when the samples were immersed in the
pressure transmission fluid (methanol/ethanol mixture), but this
shift was not discussed further. On the other hand, this
phenomenom was addressed by Wood {\it et al} in a systematic study
of the frequency of the second order overtone of the D-band
(refered as the D$^\star$ or G' band), of carbon nanotubes
dispersed in various liquids \cite {wag1}. The upshift was shown
to depend on the nature of the fluid and to increase with its
cohesive energy density (CED). Therefore, it was assigned to the
internal (van der Waals) pressure of the molecular liquids
\cite{wag1}. Radial vibrations of the carbon atoms are
particularly expected to be sensitive to this pressure. On the
other hand, under hydrostatic pressure, both the frequency of the
RBM and TM bunches is found to upshift,
 following a linear pressure dependence at low pressures of the order of 7 to 10 cm$^{-1}$.GPa$^{-1}$ (references \cite{ven} and \cite{san}, respectively).\\

We assign the upshift of the RBM from dry powders to supensions to molecular interactions between the
 surfactant molecules and the nanotubes. Since the chemical nature of the hydrophobic part of both SDS
 and TX100 is close to that of alcanes, one expects a molecular pressure of the order of the CED of
 long alcanes, {\it i.e.} of a few hundreds MPa. The shift of the RBM of the order of 5 to 6 cm$^{-1}$
 observed for EA samples (fig. 4) is of the same order of that expected for an hydrostatic pressure of
 a few hundred MPa \cite{ven}, which gives additional credit to our analysis. This analysis can also
 explain the experimental results by Rao {\it et al} \cite{rao} as well as hose by Lebedkin {\it et al} \cite{leb}.\\

{\bf B. Effect of exfoliation}\\

>From figure 3, one observes essentially no changes in the RBM frequencies of suspensions after
exfoliation. Figure 4 confirms that the upshift is globally the same for SB and for SIS. This can
appear as a surprising result since calculations predict a downshift of 10 to 15 cm$^{-1}$ for isolated
tubes with respect to infinite bundles \cite{rols,ven,hen1,rah2}. However, we point out that even though
our samples contain {\it individual} tubes, {\it i.e.} free of any intertube interaction, the nanotubes
can not be considered as {\it isolated}, since they are decorated by surfactants and immerged in a fluid.
 In our opinion, the surfactant-tube interactions explain why no shift is observed between SB and SIS.\\

The most spectacular feature of the spectra after exfoliation is the observation of new peaks at high
frequencies for SIS prepared with EA samples (marked with arrows in fig. 3a). One also observes a
significant increase of the highest frequency peaks in the spectra of HiPCO tubes (arrows in fig. 3b).
According to equation (1), these peaks correspond to SWNT of small diameter (0.8 to 0.9 nm). It is temptating
to conclude that the exfoliation process is more efficient for small diameters. However, one can not rule out
{\it a priori} an effect of some changes in the electronic properties of the tubes after exfoliation. Indeed,
calculations show that bundling makes the electronic properties vary \cite{rao,reich}. However, the nature and
amplitude of the changes is controversial. Rao {\it et al} found that intertube coupling leads to a broadening
of the van Hove singularities and to a net increase of the energy spacing for bundled nanotubes. According to
the Kataura plot, such a blueshift of the AOT favors resonance on larger tubes.
Therefore, at fixed laser energy, the RBM intensity is expected to decrease (increase) at high frequencies for
bundled (individual) tubes. From their calculations, Rao {\it et al} assigned the upshift of the RBM for
individual tubes in solution to a redshift of the AOT due to exfoliation \cite{rao}.
Part of this explanation may be right. However, we demonstrated above that the upshift in solution must be
primarily attributed to tube-liquid interactions. On the other hand, Reich {\it et al} studied the effect
of bundling on the electronic properties of SWNT through {\it ab initio} calculations \cite{reich}. They
also reported significant broadening and changes in the density of states. However, they found that the
amplitude of the changes depends on the chiral angle of the tubes. Moreover, even the nature of the change
 -redshift or blueshift- depends both on the nature of the nanotube and on the order of the transition.
 Recently, Strano {\it et al} reported changes in RBM intensities for nanotubes in various forms :
 suspensions, floculates, powders \cite{strano}... They assumed that a redshift of the AOT occurs when
 nanotubes aggregate, which allowed them to interpret well their experimental data. Unfortunately, the
 nature and strength of intertube interactions in their different
 samples were not well-characterized enough to conclude.\\

We ourselves tried to modelize the changes in the Raman spectra of
figure 3 following the same approach. Globally, the intensity of
the low-frequency peaks increases if one considers a blue-shift of
the AOT due to exfolation, in opposite to the experimental
results. On the other hand, even by considering a red-shift of the
AO, we were not able to mimic the strong increase of the high
frequency peaks except in assuming unreasonable, downshifted,
distribution of tube diameters. More important, we could never
modelize the apparition of high frequency peaks in the spectra of
EA samples. We conclude that simple electronic changes, {\it i.e.}
a general shift of the AOT, can not explain the experimental
results alone. Therefore, another phenomenom is responsible for
the increase of the RBM intensity of small tubes in the SIS
spectra. We claim that the process used to prepare SIS {\it
favorizes small tube diameters}. Such a selective effect is
supported by the high signal measured for small tubes in PL
experiments on SIS \cite{bac}. By contrast, the comparison of
optical spectra for ropes and individual tubes suggested that the
diameter distribution didn't change after exfoliation \cite{hert}.
Crossed optical and Raman measurements should be run to understand
these discrepancies. On the basis of our Raman results, we believe
that sonication is likely more effective for small diameters
because of larger curvatures and therefore a weaker van der Waals
cohesion, especially in polydisperse bundles. On the other hand,
centrifugation separates objects
 of different densities and may therefore contribute to the
diameter selection. Calculation of the diameter-dependence of the tube densities is not straightforward
since the ''thickness'' of the tubes is not known accurately. However, an experimental test of the diameter
selectivity could be achieved by performing centrifugation in solvents of various densities.\\

{\bf IV. CONCLUSION}\\

Well characterized individual tubes in suspensions were prepared from different production methods,
with different structural features. The samples were carefully characterized, especially by HRTEM and
near infrared photoluminescence, to get unambiguous signatures of the exfoliation. We studied the changes
in the Raman profiles due to dispersion of the powders in aqueous suspensions and to exfoliation into
individual tubes. Dispersion of the bundles in a liquid leads to a systematic upshift of the RBM bunches.
 This is the first important result of this study. This upshift is assigned to molecular pressure-induced
 stress. The consequences of the stress on the RBM frequencies are comparable to those of hydrostatic
 pressure. Therefore, individual suspended tubes are not relevant systems to probe the RBM shift due to
 intertube interactions in the bundles. Experiments should be run on dry deposits from the suspensions
 after dialysis or heat treatment to remove the surfactant layers.
After the tubes are exfoliated from the bundles, no frequency changes are observed in the RBM spectra.
The main feature is the strong increase of the Raman signature of small diameter tubes. This phenomenom may
be due in part to a redshift of the allowed optical transitions when tubes are exfoliated. However, it is
necessary to consider a selectivity of the individualisation process in favor of the smallest nanotubes to
explain the whole data. Such a selectivity can be due to a better effectiveness of the sonication and/or of the
centrifugation  for small diameters. In the future, a better control of the centrifugation process may be
helpful to separate the tubes as a function of their diameter.\\

{\bf ACKNOWLEDGEMENTS}\\

We acknowledge M. In, J.L. Sauvajol, S. Rols, S. Badaire and P. Poulin for fruitful discussions, A. Loiseau
and H. Amara for their help in the HRTEM
measurements, E. Doris and C. Menard for their useful advices in the preparation of the samples.
\\

\newpage

\vskip 7mm

\vskip 7mm

\vskip 7mm

\vskip 7mm

\begin{figure}[p]
\centerline{\includegraphics{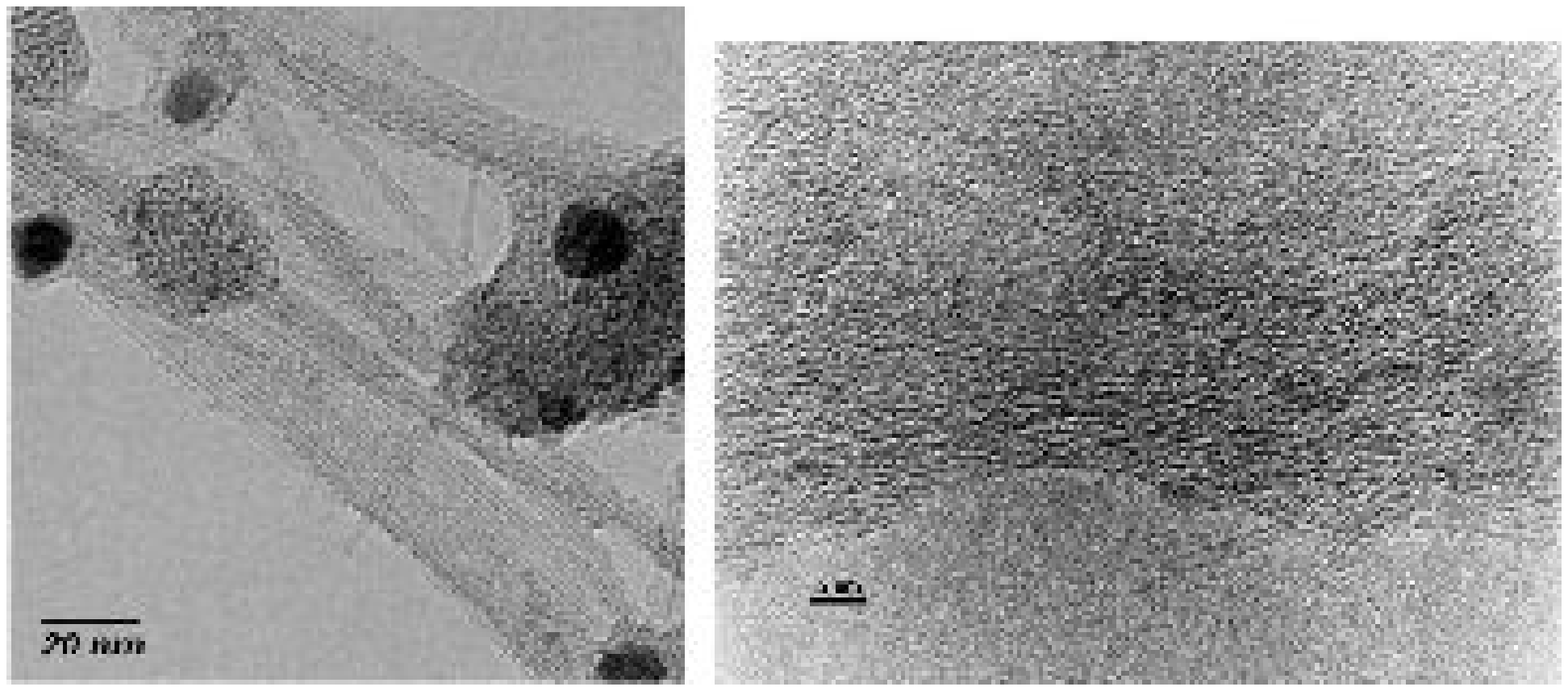}}
\caption{High resolution transmission electron microscopy from raw electric arc
samples : (a) bundles and (b) isolated tubes. See details in the text.}
\end{figure}

\begin{figure}[p]
\centerline{\includegraphics{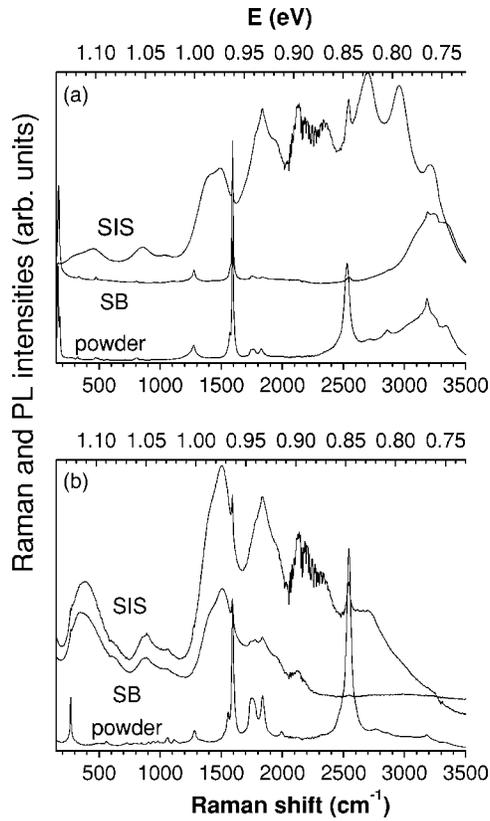}}
\caption{Raman and photoluminescence spectra for powders, suspensions of bundles and suspensions of
individual tubes, as labeled on the figures, for (a) electric arc samples and (b) HiPCO samples, excited
with a 1.16 eV laser line. The spectra were normalized and shifted along the vertical axis for clarity.
The broad and intense signal in the ranges 3200-3500 cm$^{-1}$ and 2300-2700 cm$^{-1}$ is due to OH or OD
stretchings from H$_2$O or D$_2$O, respectively. The signal fluctuations in the range 2100-2300 cm$^{-1}$
are due to absorption by atmospheric water.}
\end{figure}

\begin{figure}[p]
\centerline{\includegraphics{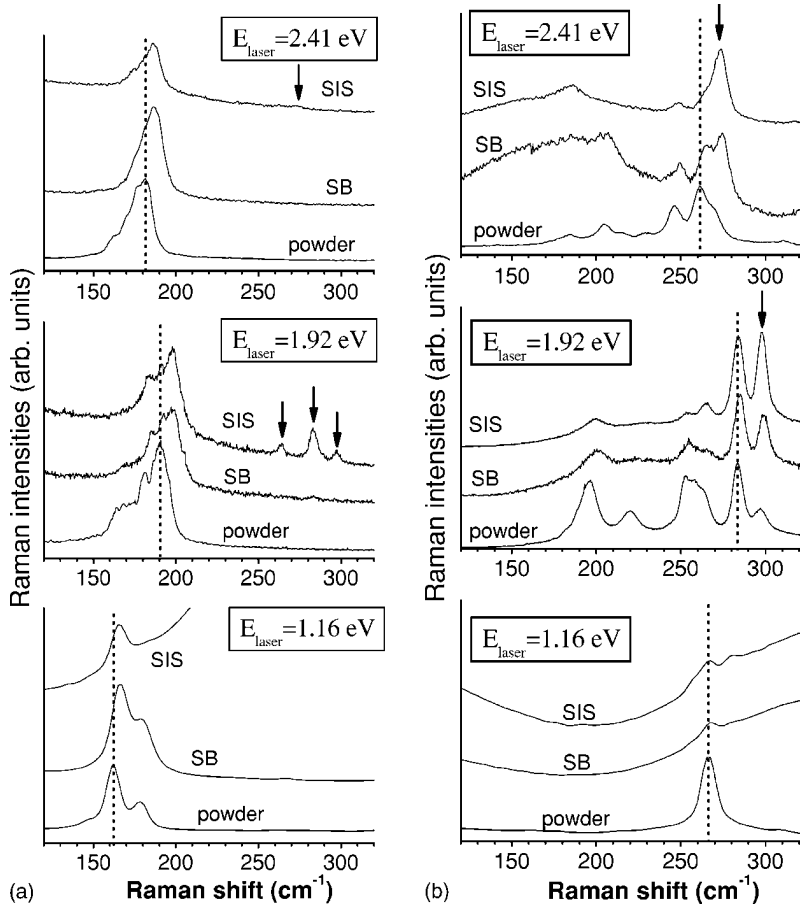}}
\caption{RBM Raman spectra for (a) electric arc samples and (b) HiPCO samples for three differents laser
lines, as labeled on the figures. The spectra were normalized and shifted along the vertical axis for
clarity. The vertical dotted lines indicate the maxima of intensity in the spectra of powders. The arrows
mark the RBM peaks whose intensity increase strongly after exfoliation (see
text).}
\end{figure}

\begin{figure}[p]
\centerline{\includegraphics{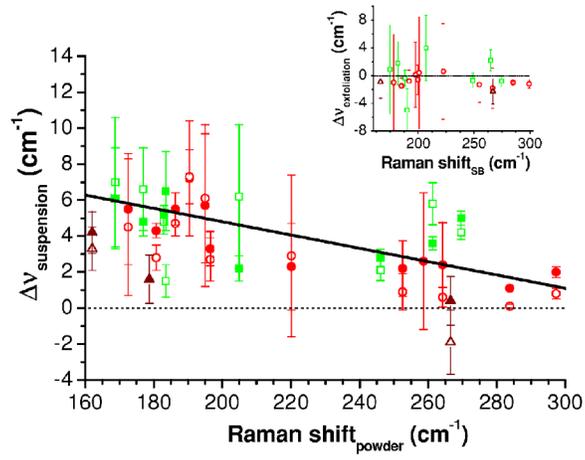}}
\caption{Shift of the RBM in the suspensions with respect to powders as a function of RBM frequency for
powders. The solid and open symbols are for SB and SIS, respectively. The squares, circles and triangles
are for laser energies 2.41, 1.92 and 1.16 eV, respectively. The error bars correspond to 95$\%$ confidence
limits. The solid line is the best linear fit for the whole data (see text). The dotted line is a guide for
the eyes.}
\end{figure}

\end{document}